%% file: tcss_lanl.tex
\documentstyle{europhys}
\input euromacr

\begin{document}
\bibliographystyle{plain}

\title{Interplay between kinetic roughening and phase ordering}

\author{Miroslav Kotrla\inst{1}
\And Milan P\v{r}edota\inst{2}}
\institute{
     \inst{1}Institute of Physics, Academy of Sciences of the Czech Republic,\\
Na Slovance 2, 180~40 Praha 8, Czech Republic 
              \\
     \inst{2}Institute of Chemical Process Fundamentals,
Academy of Sciences of the Czech Republic,\\
165 02 Praha 6, Czech Republic
}


\euro{??}{?}{1-$\infty$}{1997}
\date{\today }
\shorttitle{M. KOTRLA \etal INTERPLAY BETWEEN KINETIC ETC.}

\maketitle

\begin{abstract}
We studied interplay between kinetic roughening and phase ordering
in 1+1 dimensional  single-step solid-on-solid growth model with
two kinds of particles and Ising-like interaction.
Evolution of both geometrical and compositional properties
was  investigated
by Monte Carlo simulations for various strengths of coupling.
We found that the initial growth is strongly affected by interaction
between species, scaling exponents are enhanced and
the ordering on the surface is observed.
However, after certain time,
ordering along the surface stops and the scaling exponents cross over to
exponents of the Kardar-Parisi-Zhang universality class.
For sufficiently strong strength of coupling, ordering in vertical direction
is present and leads to columnar structure persisting for a long time.
\end{abstract}

\large

Recently there has been considerable
interest in growth induced surface roughening called
kinetic roughening \cite{review}
and in phase ordering \cite{bray94} independently.
In kinetic roughening one is interested in the evolution
of roughness
during a nonequilibrium growth process.
Phase ordering deals with the approach to equilibrium of a system quenched from a homogeneous high temperature phase
into a two-component region.
In both fields concept of scaling allowed to classify
physical processes into universality classes.
However, not very much is known so far about scaling exponents
in systems where
both processes, roughening as well as ordering,
are relevant.

Growth of many-component systems is in fact quite common situation.
Many real materials are composed of two or more components
but often the dynamics of one component is dominant and one can 
use a single-component growth model.
For example, one can consider only kinetics of Ga atoms
in studies of GaAs growth \cite{shitara92}.
We are interested here in a more complex case when dynamics of both
components is important.
It is a problem
of practical interest,
for example,  microscopic understanding of growth of alloys
is desirable \cite{hunt94b,smith96,tersoff96}.
One can study different aspects: kinetics, morphology,
scaling behaviour etc.
We are not going to model growth of any specific material but
we shall concentrate on the scaling behaviour.
The problem of growth in a system with two or more components is
interesting from pure statistical-mechanical point of view, because
growth process may belong to a new universality class
\cite{ausloos93,w_wang95,saito95};
such system might also
exhibit a nonequilibrium phase transition between the low and high
temperature region.

Let us consider a surface in a $d$-dimensional space given
by a single valued function $h({\bf r},t)$ of a $d^\prime$-dimensional
($d$$=$$d^\prime$$+$$1$) substrate coordinate ${\bf r}$.
The surface roughness is described by the surface width
$w(t,L)=\langle\sqrt{\overline{h^2}-\overline{h}^2}\rangle$,
where $t$ is the time, $L$ is the linear system size
and the bar denotes a spatial average,
$\langle ... \rangle$ a statistical average.
The surface roughness often obeys dynamical scaling law
$w(t,L)$$\propto$$L^{\zeta} f(t/L^{z})$ where the
scaling function $f(x)$ has properties: $f(x)$$=$ const., $x$$\gg$$1$
and $f(x)$$\propto$$x^{\beta}$, $x$$\ll$$1$ ($\beta$$=$$\zeta/z$).
The exponents
$\zeta$ and $z$ (or $\zeta$ and $\beta$) characterize scaling behaviour of
the surface width in a particular model and determine its universality
class \cite{review}.
This universal behaviour has been
observed in a wide variety of growth models and there has been considerable
effort in finding different possible universality classes. Many of
the growth models
studied so far (for example ballistic deposition, Eden model, restricted
solid-on-solid model, etc.) belong to the Kardar-Parisi-Zhang (KPZ)
universality class \cite{kardar86}.

There is a large menagerie of single-component growth models
which can be potentially generalized to heterogeneous case.
Moreover, there are different possible ways of generalization.
Nevertheless,
little is known so far about kinetic roughening in two-component
growth models.
This problem was probably first
considered by Ausloos {\it et al.} \cite{ausloos93}.
They introduced a generalization of the Eden model
coined as magnetic Eden model (MEM),
which contains two types of particles with probabilities of growth
given by Ising-like interaction.
Ausloos {\it et al.} found a variety of morphologies,
and they also measured the perimeter growth exponent \cite{ausloos93}.
They suggested that
the model does not belong to the KPZ universality class.
Recently Wang and Cerdeira \cite{w_wang95}
studied kinetic roughening in two 1+1 dimensional two-component growth
models with varying probability of deposition of a given
particle type. They did not found, however, a new universal behaviour.
Although the phase ordering was apparently present it was not studied
in these works.
The characteristic length in
phase ordering is a domain size $D$.
It increases with time according to a power law, $D \propto t^{\psi}$,
$\psi$ being one of the exponents
specifying a universality class.
Phase ordering is
usually a bulk process, 
but growth induced ordering
may be present just on the surface.
Then evolution of the domain size
on the surface is of interest.
The scaling in this case
has been recently studied by
Saito and M\"{u}ller-Krumbhaar \cite{saito95}
in another generalization of the Eden model (different from MEM).
They obtained $\psi =2/3$ for growth of domain size, but they did not
investigate kinetic roughening.

In this Letter
we study interplay between kinetic roughening and phase ordering.
We introduce a new growth model with two types of particles
suitable for the study of scaling
and present results of simulations in
1+1 dimensions.
Our model is based on the single-step
solid-on-solid (SOS) model, a discrete model with constraint
that the difference of heights between two neighbouring sites
is restricted to $+1$ or $-1$ only.
Advantage of this choice is that in contrast to cluster geometry,
kinetic roughening can be more easily studied and that the
single-step constraint allows simple interpretation of
the deposition process in a binary system.
We consider two types of particles,
and denote the type of a particle by a variable
$\sigma$ which assumes values $+1$ or $-1$
(not to be confused with the step size).
It is convenient to use an analogy with magnetic systems
and to consider $\sigma$ as a spin variable.
We shall use this terminology in the following, but
it should be warned that in the context of crystal growth
this may be misleading
because magnetic interactions of atoms are rather weak
and do not usually control the growth process.
One should rather think in terms of two
types of particles with different bonding energies in this case.
Therefore we shall call our model {\it two-component 
single-step} (TCSS) model.

We describe our growth model for simplicity in 1+1 dimensions but it can
be straightforwardly generalized to any dimension.
Several realization of the single-step geometry
differing by the number of
nearest neighbours for a new particle are possible.
Here we consider a variant
with three nearest neighbours which can be represented
as stacking of rectangular blocks with the height twice
the width.
During growth, particles are only added, there is neither diffusion
nor evaporation. Once a position and a type of a particle are
selected,  they are fixed forever.
Due to the single step constraint, particles can be added only at
sites with a local heigh minimum,
called growth sites.
The probability of adding a particle with a spin $%
\sigma $ to a growth site $i$
depends only on its local neighbourhood in a way analogous to rules
used in the magnetic Eden model \cite{ausloos93}.
It is proportional to $\exp (-\Delta E(i,\sigma)/k_BT)$,
where $k_B$ is Boltzmann's constant, 
$T$, and $\Delta E(i,\sigma)$ denote
thermodynamic temperature and change of energy associated with deposition
of a new particle, respectively.
The energy $\Delta E(i,\sigma)$ is given by Ising-like interaction
of a new particle with particles
on the surface within nearest neighbours
of a growth site (which are three
in the selected realization -
left, bottom and right).
When we denote the spin of a particle on the top of
a column of spins at site $i$ (surface spin) by $\sigma(i)$
then $\Delta E(i,\sigma )=-J\sigma \left[ \sigma (i-1)+\sigma (i)+\sigma
(i+1)\right] -H\sigma $.
Here $J$ is a coupling strength
and $H$ is an external field.
In the following we use for convenience
dimensionless constants $K=J/k_BT$ and $h=H/k_BT$.

We studied both geometrical and compositional
properties.
Geometry is described by the surface width $w(t,L)$ or by
height-height
correlation function
$G\left( r,t\right)$
$= \frac 1L\sum_{i=1}^L
\langle\left[ h\left( i+r,t\right)\right.$
$\left.-h\left( i,t\right) \right] ^2\rangle $.
Evolution of geometry is affected
by composition of the surface and vice versa.
Therefore we introduce quantities characterizing
composition on the surface. We concentrate only on
ordering on the surface, we do not study bulk properties here.
We consider the average domain size along the surface $D(t)$,
the spin correlation function  of surface spins
$S(r,t)=\frac 1L\sum_{i=1}^L\langle
\sigma (i+r,t)\sigma (i,t)\,\rangle $,
and magnetization $M(t)=\frac 1L\sum_{i=1}^L\langle\sigma(i,t)\rangle$.
We performed simulations for various
coupling strengths $K$ in both ferromagnetic ($K>0$)
and antiferromagnetic ($K<0$)
regimes, mostly
for zero external field $h$.
System sizes  ranged from
$L=250$ to $L=80000$, time was up to $3 \times 10^5$ monolayers (ML).
Presented results are averages over ten or more independent runs.
We start from the flat surface as usual,
but in two-component models the evolution strongly
depends on initial composition of the substrate.
We considered various possibilities (ferromagnetic, antiferromagnetic, random
composition) and finally
we have used growth on a neutral
substrate, i.e. a substrate composed of particles with zero spin,
and we let the system to evolve spontaneously
in order to avoid initial transient effects.

\begin{figure}
\caption{Time evolution of the surface domain size for
coupling strengths
$K=0.0$, $0.3$, $0.7$, $1.1$, $2.0$  and zero external field, $L=10000$.
Inset:
Saturated surface domain size as function of coupling for different system sizes
$L=250$, $L=1000$ and $L=10000$. The solid line is the fit (see text).
}
\label{fig:domain}
\end{figure}
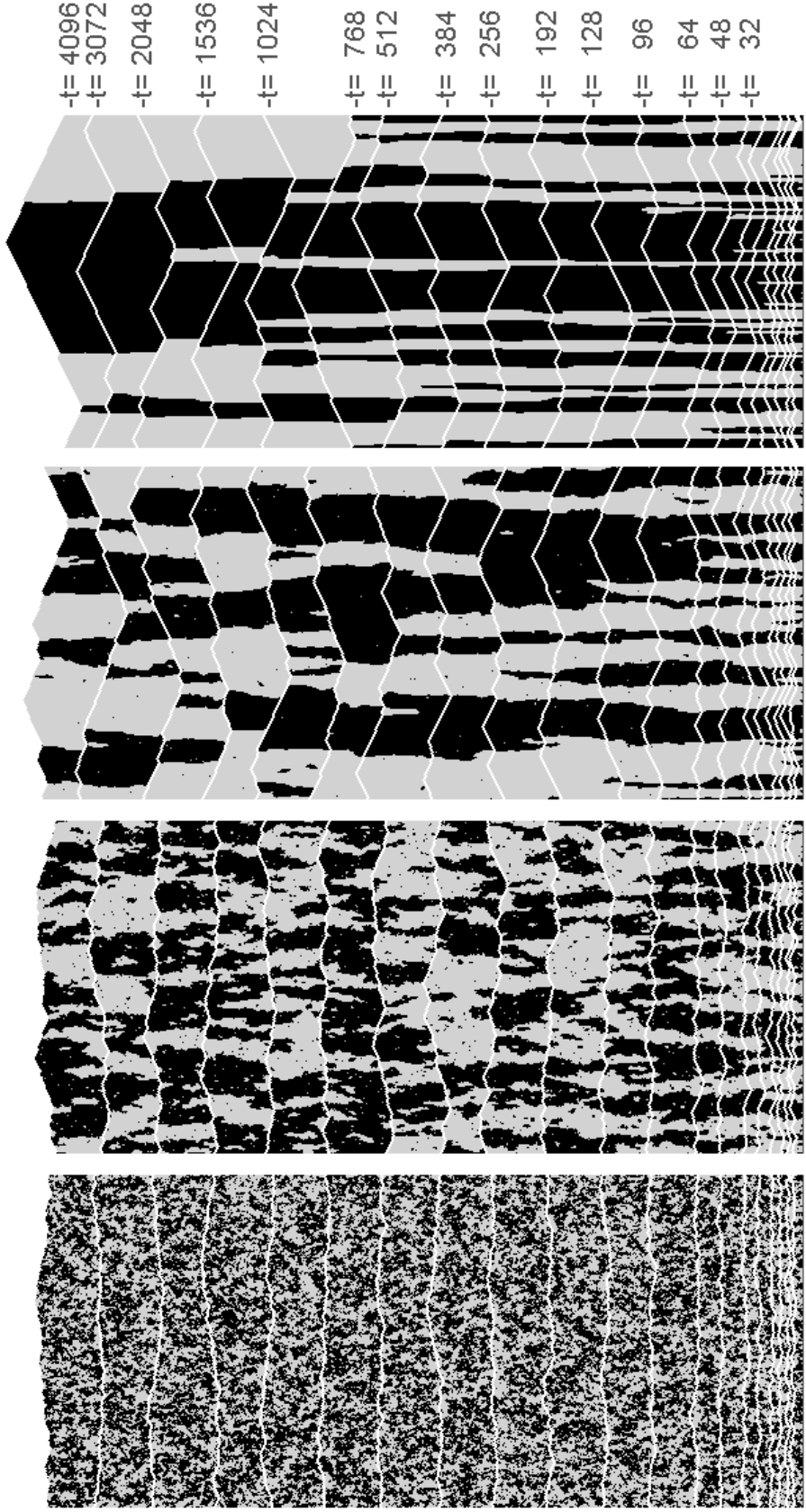
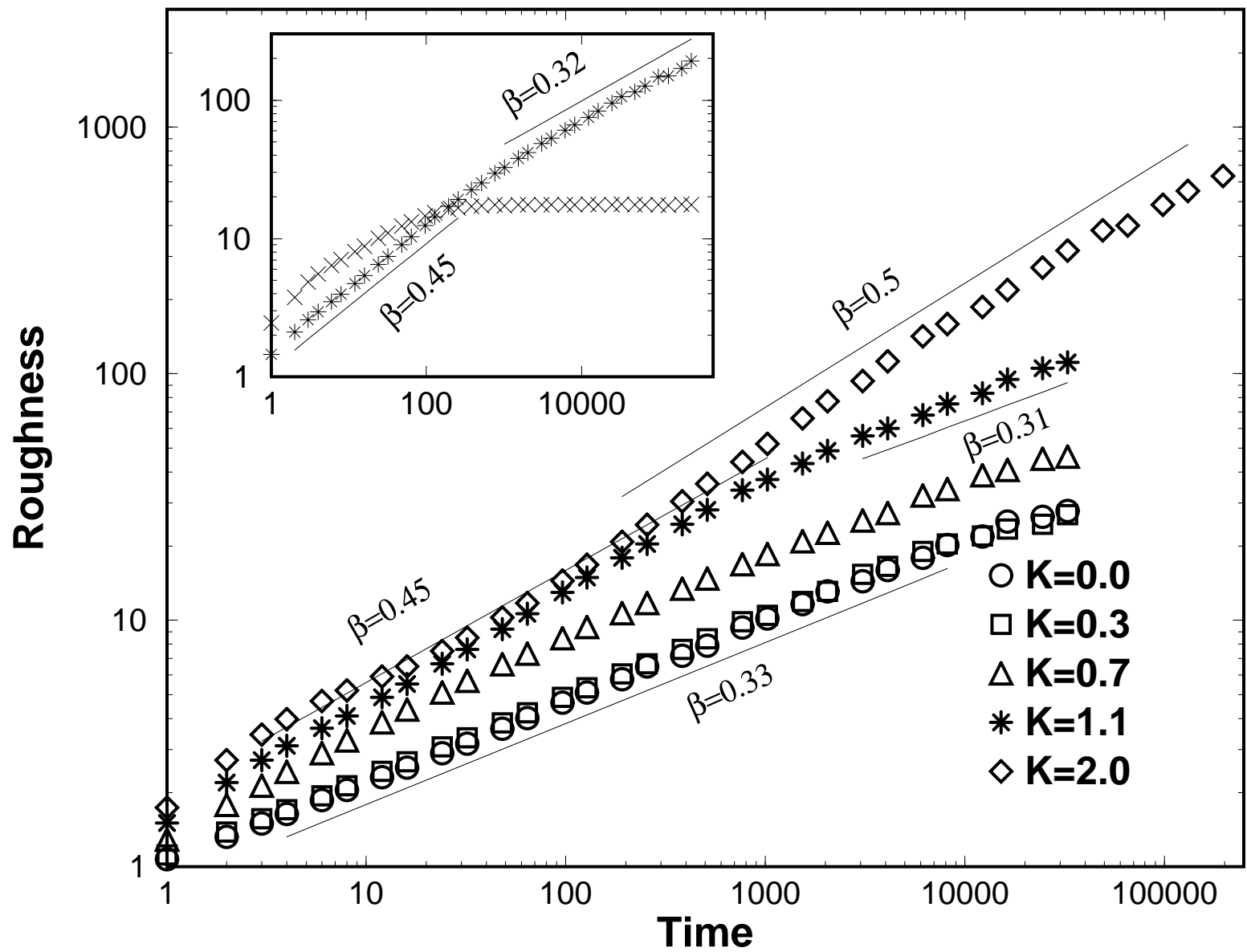
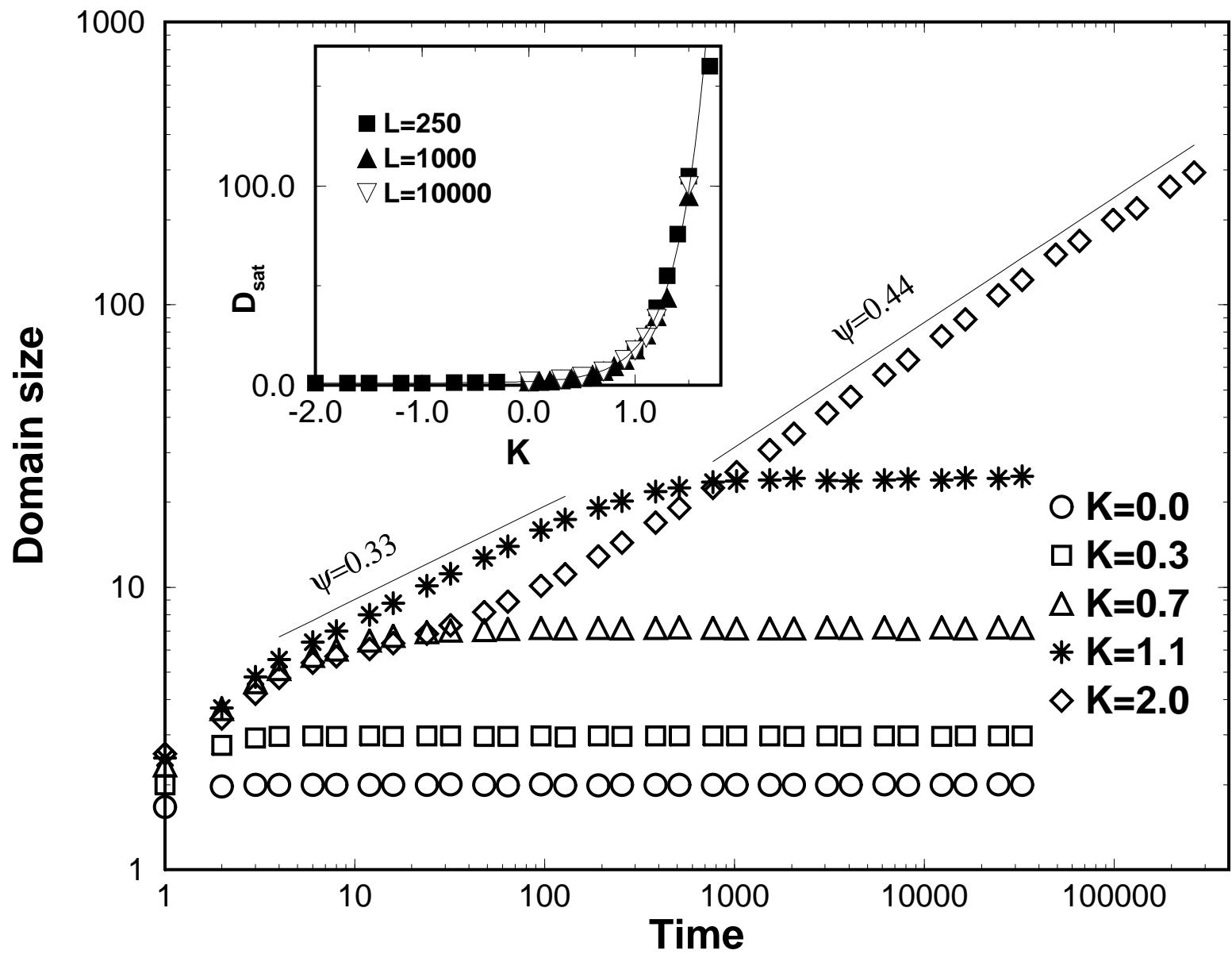

Fig. \ref{fig:conf} shows examples of time evolution
of morphologies obtained for selected couplings.
We can see that with the increasing coupling the surface becomes
more and more rough (facetted) and at the same time larger and larger
domains of the same type of surface particles are formed.
Notice also that for the larger coupling there is correlation between
domain walls and changes of the local slope, and 
a columnar structure is observed.
The time dependence of the surface width and the average
domain size  for $L=10000$
are shown in Fig. \ref{fig:rough} and Fig. \ref{fig:domain},
respectively.
When the coupling is weak evolution of the roughness
is almost the same as in the ordinary single-step model
($\beta=\beta^{(KPZ)} =1/3$).
For a larger coupling it can be seen that the
average surface width at a given time is an increasing function
of the coupling and that
ordering leads to an enhancement of the exponent $\beta_{\rm eff}$.
However, after a time $t_{\rm cross}$, that increases with the 
coupling strength, the exponent $\beta_{\rm eff}$
crosses over back to $\beta^{(KPZ)} =1/3$.
To be sure that this unexpected crossover is not a finite size
effect we performed for $K=1$ the calculation for significantly larger
system size $L=80000$
and observed the same crossover
(see inset in Fig. \ref{fig:rough}).
When the coupling is even stronger the
enhanced exponent $\beta_{\rm eff}$ is observed during the whole
simulation;
for $K=2$ we have $\beta_{\rm eff}=0.52\pm 0.02$
for times up to $3 \times 10^5$ ML.

The crossover can be understood from behaviour of the average domain size.
It increases initially according to a power law
$D\propto t^{\psi_{\rm eff}}$, $\psi_{\rm eff}$ being
an increasing function of
$K$; $\psi_{\rm eff}=0.33$ for $K=1.1$.
However, after a certain time $t^{(D)}_{\rm sat}$,  $D$ saturates
to $D_{\rm sat}(K)$.
We have checked that when the saturation is observed it
is not a finite size effect (see inset in Fig. \ref{fig:domain}),
rather  an intrinsic property of the model.
For large $K$ the domain size increases
during whole simulation, after some transient effect we observe
$\psi_{\rm eff}=0.44$ for $K=2$.
The calculation of the spin-spin correlation function, and the correlation
length $\xi^s$ derived from it,
leads to similar results:
during ordering the correlation length increases as a power law
$\xi^s(t,K)\propto t^{\kappa_{\rm eff}}$;
$\kappa_{\rm eff}\approx 1/2$
for $K=2$.
Then after a time
increasing with the
coupling ordering stops, and $\xi^s$ saturates to
$\xi^s_{\rm sat}\propto e^{\gamma K}$; $\gamma =3.25\pm 0.08$.
Both effects,  the crossover in time dependence of roughness and
the saturation of the domain size (correlation length), are apparently
related, $t_{\rm cross}\approx t^{(D)}_{\rm sat}$
(see inset in Fig. \ref{fig:rough}).

To complete the picture we need to know
the second exponent for kinetic roughening, the roughening exponent
$\zeta$.
We measured it from the spatial dependence of the
height-height correlation function
$G\left( r,t\right)$ at $t^{(w)}_{\rm sat}$ (not shown here).
The exponent has a value close to
$\zeta^{(KPZ)}=\frac{1}{2}$ for a weak coupling,
but for a larger coupling
we have found that  there is
a crossover behaviour provided
the system is sufficiently large.
We observed that the exponent $\zeta_{\rm eff}$ is
increasing with the coupling ($\zeta_{\rm eff}=0.7$ for $K=0.7$)
on distances smaller than the correlation length
$\xi^s$, and   $\zeta_{\rm eff}$ crosses over
to $\zeta^{(KPZ)}=\frac{1}{2}$ on a larger distances.
If the coupling is strong
we again do not see the crossover
but only larger exponent, $\zeta_{\rm eff}\approx1$ for $K=2$,
because the simulated system sizes are not large
enough to get into the regime $r>\xi^s$.
Exponent $\zeta=1$ is the same as in some models with surface diffusion
in which
often anomalous scaling is
observed \cite{kotrla96}.
However, this cannot be the case here since the step size
is restricted.

Our results show that for sufficiently strong coupling
the TCSS model exhibits an intermediate growth regime with 
new scaling exponents
which we estimate as $\beta =1/2$, $\zeta=1$ ($z=2$) and $\psi=1/2$.
However, these exponents are only effective and 
there is a crossover to the KPZ exponents for the medium
coupling strengths.
We estimated  the
dependence of $t^{(D)}_{\rm sat}$ ($\approx t_{\rm cross}$)
on $K$. We have found
that the saturated domain size as a function of $K$
can be well fitted in the form
$D_{\rm sat}(K)=1+\frac{1}{2}(e^{3.5K}+e^{0.7K})$
(see inset in Fig. \ref{fig:domain}).
Comparing $D_{\rm sat}\propto (t^{(D)}_{\rm sat})^{\frac{1}{2}}$
with the fit
we obtain that $t^{(D)}_{\rm sat}$ is a rapidly increasing function
of $K$, $t^{(D)}_{\rm sat}(K)\propto e^{7K}$, e.g.
$t^{(D)}_{\rm sat}(K=2) \approx 10^6$.
Due to the progressively increasing 
time and system sizes needed for simulation
we cannot exclude that there is 
a phase transition to a new true asymptotic phase at some value $K_c$.
However, we expect that the crossover in the roughness is present for
any value of $K$, but that it is hard to see it for strong coupling
because $t_{\rm cross}$ is larger than any possible simulation time.
When $K=\infty$ then the model becomes trivial:
type of a new particle added to a growth site is dictated by the majority
of particle types in the neighbourhood
and evolution of composition is fully determined by
the initial configuration.
So far all results were for zero external field.
Non-zero field leads to nonzero surface (as well as bulk)
magnetization, or in the context of alloy growth to changing stoichiometry.
We can still define the exponent $\psi$ for growth of the dominant domain
size as well as the exponents for kinetic roughening.
Effect of this symmetry breaking on the values of exponents remains to be
studied.

One would like to know if the crossover to the KPZ exponents is
generic behaviour or if facetted phase can be 
present also in the asymptotic regime.
We have also tested a different variant
of two-component growth, namely the single-step model 
which corresponds
to stacking of squares rotated by 45 degrees in which 
a new particle interacts with only two instead of three
nearest neighbours. We observed the crossover as well,
stronger coupling was needed to see enhanced exponents
and for the same coupling
the crossover time was smaller than in the variant with three 
nearest neighbours.
Is is of interest to examine kinetic roughening and phase ordering
in a different two-component growth model, in particular 
to reexamine the MEM model
in the strip geometry since in the 
cluster geometry used by Ausloos {\it et al.}
finite size effect are expected to be stronger.
Situation in higher dimensions can be investigated
by straightforward generalization.
Finally, there is question of continuous description of
the two-component growth.

In conclusion we have generalized the single-step SOS model
to two-component growth model.
It can be considered as a model for
growth of binary alloys  from a fluid containing two types of particles,
or for growth
of a colony of two kinds of bacteria etc.
The introduction of two kinds of interacting particles leads to new
phenomena, like an increase of the roughness with the increasing 
coupling strength,
larger effective scaling exponents
and ordering on the surface with corresponding pattern formation
during growth.
However, after a certain time new behaviour stops
and ordinary dynamic behaviour is restored.
Hence, growth of a crystal can be divided into two stages.
In the first,
ordering is essential and influences the evolution of geometric
properties.
In the second stage, after
saturation of domain size,
the evolution of geometric properties is similar to that of
the ordinary single-step model.
Crossover time between two regimes is
independent of system size
and is a rapidly increasing function of coupling
so that practically only the first regime may be observed.
Patterns of the grown material in saturation regime
are similar
to lamellar structure observed in eutectic growth
\cite{hunt94b}.
At the moment, it is not clear why ordering stops
and the characteristic length (given by the saturated domain size)
is selected.

\stars
 
We thank F. Slanina and N. Vandewalle for useful discussion.
This work was
supported by grant No. A 1010513 of the GA AV \v{C}R.

%

\newpage

\begin{figure}
\caption{Examples of evolution of surface profiles for
several values of coupling strength,
$K=0.3$, $0.7$, $1.1$, $2.0$  and zero external field.
Surface profiles at various times increasing as powers
are shown by white lines.
Only part of the substrate close to the surface is shown at given time,
black and grey correspond to different types of particles.
System size is $L=250$.}
\label{fig:conf}
\end{figure}

\begin{figure}
\caption{Surface width $w$ {\it vs.} time $t$ for
several values of coupling strength,
$K=0.0$, $0.3$, $0.7$, $1.1$, $2.0$  and zero external field,
$L=10000$.
Inset:
Comparison of the time dependence of the surface width and 
the domain size for $K=1.0$ and $L=80000$.
}
\label{fig:rough}
\end{figure}

\end{document}

%% file: euromacr.tex
\def\etal{{\hbox{{\tenit\ et al.\/}\tenrm :\ }}}

\def\And{{\rm and\ }}

\def\stars{\bigskip\centerline{***}\medskip}

\newif\ifboo \boofalse

\def\Review#1{\boofalse{\it #1},}
\def\Name#1{{\sc #1},}
\def\Vol#1{\ifboo Vol. {\bf #1}\else{\bf #1}\fi}
\def\Year#1{\ifboo #1\else(#1)\fi}
\def\Book#1{\bootrue{\it #1},}
\def\Page#1{\ifboo {\rm p. #1}\else{\rm #1}\fi}